\documentclass[11pt]{article}

\usepackage[T1]{fontenc}
\usepackage[utf8]{inputenc}
\usepackage{lmodern}
\usepackage{amsmath,amssymb,amsthm}
\usepackage{booktabs,array,tabularx,longtable}
\usepackage[margin=0.9in]{geometry}
\usepackage{xcolor,microtype,graphicx,float}
\usepackage{tikz}
\usetikzlibrary{positioning,arrows.meta}
\usepackage[shortlabels]{enumitem}
\usepackage{listings}
\usepackage{algorithm}
\usepackage{algpseudocode}
\usepackage[colorlinks=true,linkcolor=blue!45!black,citecolor=blue!45!black,urlcolor=blue!45!black]{hyperref}
\usepackage[nameinlink,capitalize]{cleveref}
\usepackage{parskip}

\crefname{figure}{Figure}{Figures}\Crefname{figure}{Figure}{Figures}
\crefname{table}{Table}{Tables}\Crefname{table}{Table}{Tables}
\crefname{section}{Section}{Sections}\Crefname{section}{Section}{Sections}
\crefname{appendix}{Appendix}{Appendices}\Crefname{appendix}{Appendix}{Appendices}
\crefname{listing}{Listing}{Listings}\Crefname{listing}{Listing}{Listings}
\crefname{algorithm}{Algorithm}{Algorithms}\Crefname{algorithm}{Algorithm}{Algorithms}
\crefname{proposition}{Proposition}{Propositions}\Crefname{proposition}{Proposition}{Propositions}
\crefalias{theorem}{proposition}

\definecolor{ink}{HTML}{17324D}
\definecolor{codebg}{HTML}{F5F6F6}
\definecolor{codekw}{HTML}{2F6F8F}
\definecolor{codestr}{HTML}{A45542}
\definecolor{codecom}{HTML}{4F8F70}

\lstdefinestyle{py}{
  language=Python, basicstyle=\ttfamily\small,
  keywordstyle=\color{codekw}\bfseries, stringstyle=\color{codestr},
  commentstyle=\color{codecom}\itshape, backgroundcolor=\color{codebg},
  frame=single, rulecolor=\color{codebg}, framerule=0pt, showstringspaces=false,
  breaklines=true, columns=fullflexible, captionpos=b,
  aboveskip=6pt, belowskip=4pt, xleftmargin=8pt, xrightmargin=8pt,
}
\lstdefinestyle{sh}{
  language=bash, basicstyle=\ttfamily\small,
  keywordstyle=\color{codekw}\bfseries, stringstyle=\color{codestr},
  commentstyle=\color{codecom}\itshape, backgroundcolor=\color{codebg},
  frame=single, rulecolor=\color{codebg}, framerule=0pt, showstringspaces=false,
  breaklines=true, columns=fullflexible, captionpos=b,
  aboveskip=6pt, belowskip=4pt, xleftmargin=8pt, xrightmargin=8pt,
}
\algrenewcommand\algorithmicrequire{\textbf{Input:}}
\algrenewcommand\algorithmicensure{\textbf{Output:}}

\newtheorem{theorem}{Theorem}[section]
\newtheorem{proposition}[theorem]{Proposition}

\theoremstyle{definition}

\newcommand{\D}{\mathbb D}

\newcommand{\Q}{\mathbb Q}

\newcommand{\Sstar}[1]{\mathcal S^*(#1)}
\newcommand{\pkg}{\texttt{geometric\allowbreak-function\allowbreak-atlas}}
\newcommand{\pkgversion}{0.2.1}
\newcommand{\cli}{\texttt{gfa}}
\newcommand{\sub}{\prec}
\newcommand{\out}[1]{\textsc{#1}}

\hypersetup{
  pdftitle={Geometric Function Atlas: certified computing for geometric function theory in Python},
  pdfauthor={Prasanna Devadiga, Kishan Gurumurthy, Arya Suneesh, Pushparaj Devadiga, Asha Sebastian}
}

\title{\textcolor{ink}{Geometric Function Atlas: certified computing for\\
geometric function theory in Python}}
\author{Kishan Gurumurthy\footnotemark[1]\textsuperscript{,\ 1}, %
        Pushparaj Devadiga\footnotemark[1]\textsuperscript{,\ 2}, %
        Prasanna Devadiga\thanks{These authors contributed equally.}\textsuperscript{,\ 1}, %
        Arya Suneesh\footnotemark[1]\textsuperscript{,\ 1},\\
        Asha Sebastian\textsuperscript{1}\\[6pt]
\normalsize \textsuperscript{1}Department of Computer Science and Engineering,\\
\normalsize Indian Institute of Information Technology Kottayam, Kerala, India\\[4pt]
\normalsize \textsuperscript{2}K.~J. Somaiya College of Engineering, Mumbai, India\\[4pt]
\normalsize \texttt{kishangurumurthy@outlook.com}, \texttt{pdevadiga451@gmail.com}  \\
\texttt{devadigaprasanna28@gmail.com}, %
\normalsize\texttt{aryasuneesh3@gmail.com}, %
\texttt{asha@iiitkottayam.ac.in}}
\date{September 2026}

\begin{document}
\maketitle

\begin{abstract}
We describe \pkg, our open-source Python package for the sharp extremal
problems of geometric function theory. We organise it around a catalogue
of thirty-nine Ma--Minda starlike generators. From this catalogue we
compute exact Taylor coefficients, closed-form Fekete--Szeg\H{o}
constants, exact coefficients of the Ma--Minda extremal function, and
admissibility screens. Our verifier answers membership questions for
normalised polynomials at three levels of evidence: a floating-point grid
screen, an exact sufficient condition decided in rational arithmetic, and
a certified interval enclosure at the worst screened point. Every answer
names the level at which we obtained it. We ship a checksummed artifact
snapshot with three hundred and six coefficient certificates and seven
hundred and two directed inclusion radii. Eight reviewed radius lanes
carry certificates whose proof chains we replay symbolically, and we
re-execute every coefficient certificate through our exact Schur-parameter
machinery on request. We emit all results through one versioned envelope
that records the method, the evidence status, the assumptions, and the
artifact identifiers. Two optional laboratories apply the same discipline
to cryptographic S-box metrics and to image-quality metrics. We present
our design, state as propositions what each tier establishes, follow one
radius lane from screen to replayed certificate, report measured timings,
and place our package among symbolic-algebra, rigorous-numerics, and
mathematical-database software. We release \pkg{} under the MIT licence
on the Python Package Index and at
\url{https://github.com/Prasanna28Devadiga/geometric-function-atlas}.
\end{abstract}



\section{Introduction}\label{sec:intro}

The extremal problems of geometric function theory ask for sharp
constants. Let
\[
  f(z) \;=\; z + a_2 z^2 + a_3 z^3 + \cdots
\]
be analytic and normalised on the unit disk $\D$. One asks for the largest
value of a functional of the Taylor coefficients of $f$ over a given
class, or for the largest disk on which one geometrically defined class
lies inside another. The two founding examples are Bieberbach's
coefficient estimate~\cite{Bieberbach,DeBranges,Duren} and the
Fekete--Szeg\H{o} inequality~\cite{FeketeSzego,KeoghMerkes}. Later work
asked the same kind of question for Hankel
determinants~\cite{LeeRavichandran}, for Zalcman
functionals~\cite{ZalcmanMa,RavichandranVerma}, and for a growing list of
inclusion radii between generator-defined classes. In every case the goal
is a sharp constant together with an extremal function that attains it.

Since the work of Ma and Minda~\cite{MaMinda}, most of this activity has
been organised around a single analytic generator $\varphi$. A generator
satisfies $\varphi(0) = 1$, $\varphi'(0) > 0$ and $\Re\varphi(z) > 0$ on
$\D$, and its image $\varphi(\D)$ is starlike with respect to $1$ and
symmetric about the real axis. The class $\Sstar\varphi$ consists of the
normalised $f$ for which $zf'(z)/f(z)$ is subordinate to $\varphi$. The
subordination expresses the low-order coefficient functionals of $f$
through the first coefficients $B_1, B_2, B_3$ of $\varphi$. Over the past
decade many generators have been studied in this framework: the
exponential, sine, cardioid, lemniscate, nephroid, tangent, lima\c{c}on,
hyperbolic-cosine, petal, bean, Bell and modified-sigmoid generators, and
the classical Janowski and parabolic families. Each new $\varphi$ raises
the same questions again. The sharp constants are published class by
class, with different normalisations and different conventions for
sharpness, and the resulting literature is hard to compare by hand.

A researcher meets these constants in three settings, which match the
stages of experimental mathematics described by Borwein and
Bailey~\cite{BorweinBailey,BorweinBaileyNotices}. In exploration, one
computes coefficients and functionals quickly and at modest precision, to
see whether a proposed reduction is promising. In proof, one works
symbolically with the Schur-parameter representation of the coefficient
body~\cite{Schur,Simon}, and for radius problems with the boundary of the
target domain, in the hope of a closed form. In verification, of one's own
result or of a published value, one wants a machine-checkable record that
the constant is correct, sharp and attained. The three settings accept
different amounts of slack. A numerical bracket is welcome in
exploration, is handled with care in proof, and does not count as
verification.

We built \pkg{} around this distinction. Its basic objects are those of
the field: Ma--Minda generators, canonical class labels, coefficient
functionals and directed radius records. Every result we return names the
kind of evidence behind it. Where a quantity can be computed exactly, we
return the exact value. Where we can offer only a numerical screen, we
label it a screen. Where we can offer a certified enclosure at a point, we
label it an enclosure. This matches the practice of our field, where a
bracket presented as a sharp constant is an error of category, and a
sharp result without an attaining extremal function is incomplete.

We develop the mathematics behind our package in our companion
paper~\cite{AtlasPaper}. There we introduce our registry, prove the sharp
radius results, and derive the coefficient theorems whose certificates our
package ships. That content consists of a catalogue of thirty-nine
generators, a set of reviewed exact inclusion radii, and the certified
coefficient bounds that go with them. In this paper we describe the
software. We explain its architecture and public interface, state as
propositions what each operation establishes, and show the
reproducibility tools we offer to a referee. We produced every number,
printed output and timing quoted below with the released package, version
\pkgversion, on the hardware described in \cref{sec:perf}. The scripts
that produced them accompany the manuscript.

\paragraph{Scope.}
We address the extremal problems that have a Ma--Minda formulation. These
are the Fekete--Szeg\H{o} functional and our catalogue of certified
coefficient bounds; membership of normalised polynomials in the starlike
and convex classes and under the Becker and Nehari univalence criteria;
and directed inclusion radii between generator-defined starlike classes.
Meromorphic classes, multivalent functions, close-to-convex families
outside the Ma--Minda framework, and quasiconformal extensions are
outside the present release. We have planned several of these extensions
and return to them in \cref{sec:future}.

\paragraph{Organisation.}
\Cref{sec:overview} states our design principles and main capabilities.
\Cref{sec:install} covers installation and a short tour.
\Cref{sec:features} treats the mathematical features one by one, giving
for each the mathematical object, the algorithm we run, and an example
with its output. \Cref{sec:guarantees} states as propositions what each
operation establishes. \Cref{sec:case} follows one radius lane from
screen to replayed certificate. \Cref{sec:perf} reports timings and
compares our replay with a hand-written check. \Cref{sec:repro} describes
our reproducibility architecture and \cref{sec:related} the relation to
existing software. \Cref{sec:conclusion} concludes, and \cref{sec:future}
lists the next steps. \Cref{app:labs} describes the two application
laboratories.

\section{Overview and design principles}\label{sec:overview}

\begin{figure}[H]
\centering
\begin{tikzpicture}[x=1mm, y=1mm,
  box/.style={draw=ink, rounded corners=2pt, align=center, font=\footnotesize,
              inner sep=3pt, text width=26mm, minimum height=14mm, fill=white},
  store/.style={box, text width=48mm, minimum height=12mm, fill=codebg},
  arr/.style={-{Latex[length=1.8mm]}, thick, ink}
]
\node[store] (cat)  at (-33.5, 0) {Generator catalogue\\ thirty-nine Ma--Minda generators};
\node[store] (snap) at (50.25, 0) {Artifact snapshot\\ 306 coefficient certificates, 702 directed radii, 8 proof chains};
\node[box] (exact)   at (-67, -24)  {Exact core\\ series, Fekete--Szeg\H{o}, extremal coefficients};
\node[box] (screens) at (-33.5, -24) {Screens\\ admissibility, membership, containment};
\node[box] (verify)  at (0, -24)    {Tiered verifier\\ screen, symbolic, rigorous; witnesses};
\node[box] (replay)  at (33.5, -24) {Certificate replay\\ Schur re-evaluation, radius chains};
\node[box] (recon)   at (67, -24)   {Reconciliation\\ literature status, open problems};
\node[box, fill=ink!8, text width=150mm, minimum height=10mm] (env) at (0, -44)
  {Result envelope: canonical inputs, method, evidence status, assumptions, artifact identifiers};
\draw[arr] (cat) -- (exact);
\draw[arr] (cat) -- (screens);
\draw[arr] (cat) -- (verify);
\draw[arr] (snap) -- (replay);
\draw[arr] (snap) -- (recon);
\foreach \n in {exact, screens, verify, replay, recon}
  \draw[arr] (\n) -- (env.north -| \n);
\end{tikzpicture}
\caption{Architecture of the package. The catalogue feeds the exact core,
the screens and the tiered verifier. The artifact snapshot feeds the
certificate replays and the reconciliation table. Every operation returns
the same result envelope.}
\label{fig:arch}
\end{figure}
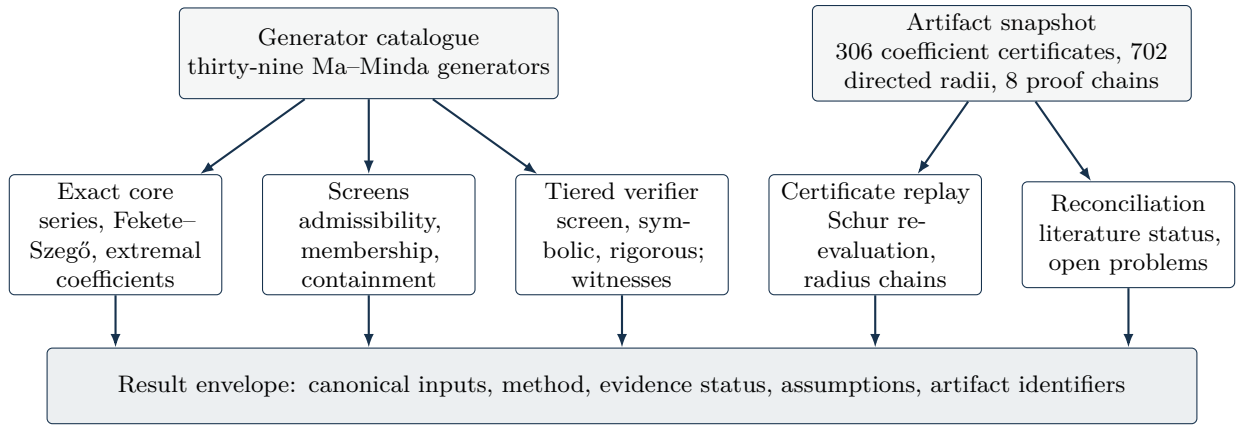

Our package has a small Python interface and a command-line tool, \cli.
With it we ship a checksummed artifact snapshot, identified as
\texttt{2026.08.11}. The snapshot contains the thirty-nine generator
definitions, three hundred and six coefficient certificates, the
Schur-parameter expansions of the low-order coefficients for twenty-two
classes, a table of open problems, a literature-reconciliation table, and
a directed radius snapshot of seven hundred and two ordered pairs. Eight
of those pairs carry replayable proof chains. The release we describe is
version \pkgversion, under the MIT licence. \Cref{fig:arch} shows how the
parts fit together.

Four principles shaped our design. Each one turns a familiar habit of the
field into a rule of the software, and we state them here so that the
reader can recognise them in the interface that follows.

\paragraph{Exact wherever the computation is exact.}
We compute Taylor coefficients of a generator, Fekete--Szeg\H{o}
constants, and the coefficients of the Ma--Minda extremal function in
exact rational or algebraic arithmetic. We return them as SymPy
expressions, so they can be used in further exact computation.
Coefficient functionals beyond the Fekete--Szeg\H{o} family we ship as
versioned certificates, and on request we re-execute the extremal stored
in each certificate through our exact Schur-parameter machinery. We
compute a decimal only when the caller asks for one.

\paragraph{Every result names its evidence.}
Every result carries an evidence label from a fixed list. One label,
\emph{proven exact under declared assumptions}, marks an exact
computation or a certificate replay that passed every required check. A
second, \emph{certified enclosure}, marks an interval enclosure at a
point. A third, \emph{numerical screen}, marks a floating-point grid
evaluation. A fourth, \emph{unresolved}, marks a claim that the available
evidence cannot settle. Our membership verifier also reports the tier
requested: screen, symbolic or rigorous. These labels keep a screen from
being read as a proof, and an enclosure from being read as a sharp
constant. We record literature status in a separate field, so that
computational evidence and bibliographic assessment stay apart.

\paragraph{Reference data is versioned and checksummed.}
We read certificates, expansions, bounds, reconciliation rows and radius
records from a read-only snapshot. A manifest lists its eight files with
their SHA-256 digests. One call returns the snapshot identity and another
recomputes the digests (\cref{lst:artifacts}). The radius snapshot has
its own identifier, \texttt{gft-radius-snapshot:2026.08.09}, together
with the commit of our research repository from which we baked it. We
stamp these identifiers into every result envelope, so a constant
reported as ``\pkg{} v\pkgversion, artifact 2026.08.11'' names one exact
bit sequence.

\paragraph{A stable result contract.}
Every result is a record with the same fixed set of named fields, so a
script can read any result in the same way. The fields fall into four
groups. The first says what was computed: the kind of result, its
canonical inputs, the exact expressions, and a small graph that encodes
each exact expression; the graph is the reference form, and the printed
strings are for display. The second says how it was established: the
method, the evidence status, the assumptions, and a verification report
that lists each check by name with its expected and observed values. The
third says where the data came from: the source references and the
versions of the package and of the artifacts. The fourth says how the
result stands: its literature status, a novelty flag, its failure state,
and the schema version. The Python functions and the command-line tool
return the same record. When an operation cannot complete, it returns one
of five failure states instead, namely invalid input, unsupported,
unresolved, resource limit and corrupt artifact, which the command-line
tool maps to the exit codes $2$ to $6$. The full contract is documented
in the source distribution.

These four principles let one interface serve the three settings of the
Introduction. A researcher can explore a new generator, a referee can
recompute a published constant, and a downstream application can consume
labelled numbers.

\section{Installation and quick tour}\label{sec:install}

Our package requires Python 3.10 or later and installs from the Python
Package Index. The route we support uses the \texttt{uv} tool
manager,\footnote{\url{https://docs.astral.sh/uv/}} which provisions
Python 3.12 automatically on macOS, Linux and Windows. A plain
\texttt{pip} installation works as well:
\begin{lstlisting}[style=sh,caption={Installation with the optional laboratories; either route gives the \texttt{gfa} command.},label={lst:install-uv}]
  curl -LsSf https://astral.sh/uv/install.sh | sh
  uv tool install 'geometric-function-atlas[lab]' --python 3.12
  pip install 'geometric-function-atlas[lab]'     # alternative, inside your own environment
  gfa --version            # geometric-function-atlas 0.2.1
  \end{lstlisting}

The bracket suffix \texttt{[lab]} adds NumPy and enables the laboratories
of \cref{app:labs}. The command \texttt{gfa walkthrough} prints a short
guided tour. The session below runs four of the operations described in
this paper and shows selected lines of their output:

\begin{lstlisting}[style=sh,caption={A first session at the command line.},label={lst:cli-tour}]
gfa coefficients exponential --order 4
#   exact_expressions: {'coefficients': ['1', '1/2', '1/6', '1/24'], 'generator': 'exp(z)'}
#   evidence_status: proven_exact_under_declared_assumptions

gfa fekete-szego exponential --mu 0
#   exact_expressions: {'B1': '1', 'B2': '1/2', 'generator': 'exp(z)', 'value': '3/4'}

gfa verify --coefficients "1" --property starlike --max-cost rigorous
#   Outcome: CERTIFIED CRITERION VIOLATION (rigorous)
#   Worst screened point: z = -0.98 1.20015e-16i

gfa verify-radius-certificate sine sigmoid
#   PROVEN: sine->sigmoid
#     PASS psi composition reduces to the logarithmic sine form
#     PASS d/dz atanh(sin z) = sec z
#     PASS |cos(x+iy)|^2 = cos^2 x + sinh^2 y
#     PASS angular-bound remainder is sinh(b)^2 >= 0
#     PASS 2*atanh(sin(r*)) = 1
\end{lstlisting}

Every result-printing command accepts \texttt{--json} and then emits the
envelope of \cref{sec:overview}. We refuse malformed input with exit code
$2$. For instance we reject \texttt{--mu 0.5}, because $\mu$ must be an
integer or an exact fraction such as \texttt{1/2}. The same session in
Python is:

\begin{lstlisting}[caption={The same session in Python.},label={lst:py-tour}]
from geometric_function_atlas import (
    generator_series, fekete_szego, verify_function, verify_radius_certificate,
)

series = generator_series("exponential", order=4)
print(series.coefficients)                  # (1, 1/2, 1/6, 1/24)

fs = fekete_szego("exponential", mu=0)
print(fs.value, fs.evidence_status)         # 3/4 proven_exact_under_declared_assumptions

rig = verify_function([1.0], property="starlike", max_cost="rigorous")
print(rig.outcome, rig.evidence_kind)       # certified_violation certified_enclosure

rep = verify_radius_certificate("sine", "sigmoid")
print(rep.status, rep.certified)            # proven True
\end{lstlisting}

The Python functions are our primary interface; the command-line tool
calls them. We ran every example in the sections that follow against a
default installation of version \pkgversion, and the printed comments
show its output.

\section{Mathematical features}\label{sec:features}

We organise the interface along the natural divisions of the field. For
each capability we state the mathematical object, give the algorithm we
run, and show a short example. We begin with the generators, because we
derive every other object in our package from them.

\subsection{Ma--Minda generators and coefficient series}\label{sec:generators}

Recall that a Ma--Minda generator is an analytic function
\begin{equation}\label{eq:generator}
  \varphi(z) \;=\; 1 + B_1 z + B_2 z^2 + B_3 z^3 + \cdots,
  \qquad
  \varphi(0) = 1,\quad B_1 = \varphi'(0) > 0,\quad \Re\varphi > 0
  \text{ on } \D,
\end{equation}
whose image $\varphi(\D)$ is starlike with respect to $1$ and symmetric
about the real axis. Subordination of $zf'(z)/f(z)$ to $\varphi$ defines
the class $\Sstar\varphi$~\cite{MaMinda}, and the sharp constants of the
class depend on the first coefficients $B_1, B_2, B_3$. We therefore store
each registered $\varphi$ as a record with four fields: a canonical key, a
display name, the exact SymPy expression in the reserved symbol $z$, and a
citation to the paper that introduced it. We compute the Taylor
coefficients by exact series expansion, as in \cref{alg:series}. The
result carries a verification report that records $\varphi(0) = 1$ and
that the returned list agrees with an independent re-expansion.

\begin{algorithm}[H]
\caption{\textsc{GeneratorSeries}$(\varphi, N)$}\label{alg:series}
\begin{algorithmic}[1]
\Require generator $\varphi$ as an exact expression in $z$; order $N \ge 1$
\Ensure exact coefficients $(B_1, \dots, B_N)$ of $\varphi(z) = 1 + \sum_{n\ge1} B_n z^n$
\State $T \gets$ Taylor polynomial of $\varphi$ at $z = 0$ through degree $N$, in exact arithmetic
\State \Return $(\,[z^1]T,\ \dots,\ [z^N]T\,)$, each an element of $\Q$ or of an algebraic extension of $\Q$
\end{algorithmic}
\end{algorithm}

\begin{lstlisting}[caption={Enumerating, expanding, and constructing generators.},label={lst:generators}]
from geometric_function_atlas import Generator, generator_series, list_generators, z

gens = list_generators()
print(len(gens), gens[11].key, gens[11].expression)   # 39 exponential exp(z)

print(generator_series("exponential", order=4).coefficients)   # (1, 1/2, 1/6, 1/24)
print(generator_series("sine", order=4).coefficients)          # (1, 0, -1/6, 0)

# A user-supplied generator: phi(z) = 1 + z/2, the Janowski class S*[1/2, 0].
half = Generator(key="half", name="Janowski S*[1/2, 0]",
                 expression=1 + z/2, citation="User supplied")
print(generator_series(half, order=3).coefficients)            # (1/2, 0, 0)
\end{lstlisting}

The thirty-nine generators in our catalogue cover the
exponential~\cite{Mendiratta}, sine~\cite{ChoSine},
cardioid~\cite{SharmaCardioid}, lemniscate~\cite{SokolStankiewicz},
lima\c{c}on~\cite{MasihKanas}, nephroid~\cite{WaniNephroid},
hyperbolic-cosine~\cite{MundaliaKumar}, tangent~\cite{UllahTanh},
petal~\cite{KumarArora}, bean~\cite{KumarYadav}, Bell~\cite{BellClass}
and modified-sigmoid~\cite{GoelKumar} families. They also include the
classical Janowski~\cite{Janowski} and parabolic~\cite{Ronning} families,
the classes of order $\alpha$ and the strongly starlike classes at
several parameter values, and further classes from the recent
literature. Our generator constructor also accepts a user-supplied
expression, so a $\varphi$ that arises during a proof can be used through
the same interface as the catalogue. The constructor checks that the
expression is an exact SymPy object in the reserved symbol $z$ alone.
Analytic admissibility of a user-supplied generator, that is, the
conditions of \eqref{eq:generator}, remains an assumption declared by the
caller, and we record it in the result envelope as caller-supplied
provenance. For the example above the conditions are immediate:
$\Re(1 + z/2) \ge 1/2$ on $\D$, the coefficients are real, and
$z\varphi'(z)/(\varphi(z) - 1) \equiv 1$.

The coefficients returned by \cref{alg:series} are exact rational or
algebraic numbers. The reduction in the next subsection needs this
exactness. It depends on the ratio $B_2/B_1$ and on $B_1^2$, and a
floating-point $B_2$ would carry a rounding error into the
Fekete--Szeg\H{o} constant, which would then lose its sharpness.

\subsection{Fekete--Szeg\H{o} constants}\label{sec:fs}

Fix $\Sstar\varphi$ and a rational parameter $\mu$, and consider the
Fekete--Szeg\H{o} functional $a_3 - \mu a_2^2$. Write
$zf'(z)/f(z) = \varphi(\omega(z))$ with a Schwarz function
$\omega(z) = c_1 z + c_2 z^2 + \cdots$. Comparing coefficients gives
$a_2 = B_1 c_1$ and $2a_3 - a_2^2 = B_1 c_2 + B_2 c_1^2$, so
\begin{equation}\label{eq:fs-reduction}
  a_3 - \mu\, a_2^2
  \;=\;
  \frac{B_1}{2}\Bigl[\, c_2 + \nu\, c_1^2 \,\Bigr],
  \qquad
  \nu \;=\; \frac{B_2}{B_1} + (1 - 2\mu)\, B_1 .
\end{equation}
The Keogh--Merkes inequality~\cite{KeoghMerkes} for Schwarz functions
gives $|c_2 + \nu c_1^2| \le \max\{1, |\nu|\}$, with equality for
$\omega(z) = z^2$ when $|\nu| \le 1$ and for a rotation of
$\omega(z) = z$ when $|\nu| \ge 1$. Substituting into
\eqref{eq:fs-reduction} gives the sharp bound
\begin{equation}\label{eq:fs-bound}
  \max_{f \in \Sstar\varphi} |a_3 - \mu\, a_2^2|
  \;=\;
  \frac{B_1}{2}\, \max\{\, 1,\; |\nu| \,\},
\end{equation}
which is piecewise in $\mu$ and changes branch where $|\nu| = 1$. We
evaluate \eqref{eq:fs-bound} exactly, as in \cref{alg:fs}, after
checking the two conditions we can decide: $B_1$ is real and positive,
and $B_2$ is real.

\begin{algorithm}[H]
\caption{\textsc{FeketeSzego}$(\varphi, \mu)$}\label{alg:fs}
\begin{algorithmic}[1]
\Require generator $\varphi$; parameter $\mu \in \Q$ given as an integer, an exact fraction, or a SymPy rational
\Ensure exact value of $\max_{f\in\Sstar\varphi}|a_3 - \mu a_2^2|$ with a five-check verification report
\State $(B_1, B_2) \gets$ \textsc{GeneratorSeries}$(\varphi, 2)$
\If{$B_1$ is not real and positive, or $B_2$ is not real} \textbf{fail} with the state \out{invalid input} \EndIf
\State $\nu \gets B_2/B_1 + (1-2\mu)\,B_1$ \Comment{exact arithmetic}
\State $v \gets \tfrac{B_1}{2}\max\{1, |\nu|\}$, simplified
\State record checks: $\varphi(0)=1$; $(B_1,B_2)$ agree with re-expansion; $B_1>0$ real; $B_2$ real; $v$ equals the closed form
\State \Return $(v, B_1, B_2, \mu, \text{report})$
\end{algorithmic}
\end{algorithm}

For the exponential generator $\varphi(z) = e^z$ we have $B_1 = 1$ and
$B_2 = 1/2$, so $\nu = 3/2 - 2\mu$. The bound \eqref{eq:fs-bound} equals
$\tfrac12\cdot\tfrac32 = \tfrac34$ at $\mu = 0$, equals $\tfrac12$ for
$\tfrac14 \le \mu \le \tfrac54$ where $|\nu| \le 1$, and equals
$\tfrac12\cdot\tfrac52 = \tfrac54$ at $\mu = 2$. We return exactly these
values:

\begin{lstlisting}[caption={Fekete--Szeg\H{o} constants for the exponential class and for a user-supplied class.},label={lst:fs}]
from geometric_function_atlas import fekete_szego

for mu in (0, "1/4", 1, "5/4", 2):
    print(mu, fekete_szego("exponential", mu=mu).value, end=" | ")
print()
# 0 3/4 | 1/4 1/2 | 1 1/2 | 5/4 1/2 | 2 5/4 |

r = fekete_szego("exponential", mu=0)
print(r.b1, r.b2, r.decimal(precision=5))    # 1 1/2 0.75000
print(r.evidence_status)                     # proven_exact_under_declared_assumptions

print(fekete_szego(half, mu=0).value)        # 1/4
\end{lstlisting}

For the user-supplied generator $1 + z/2$ of \cref{lst:generators} we
have $B_1 = 1/2$ and $B_2 = 0$, so $\nu = 1/2$ and the bound is $1/4$.
We accept $\mu$ as an integer, as an exact fraction written
\texttt{"p/q"}, or as a SymPy rational. We refuse decimal input, because
a decimal would replace the exact transition points of
\eqref{eq:fs-bound} by floating-point approximations. The returned value
is a SymPy expression, so the user can use it in further computation,
compare it exactly with a published constant, or request a decimal to
any precision. The envelope records the two Ma--Minda conditions we
checked and lists analytic admissibility of $\varphi$ among its declared
assumptions.

\subsection{Class operations}\label{sec:classes}

With the generator fixed, we turn to the class $\Sstar\varphi$ itself. We
provide four operations, and they divide by the kind of evidence they
produce. The extremal coefficients are exact. The admissibility check
decides two conditions exactly and screens three more on a sampled grid.
Membership and containment are screens by winding number against a
sampled boundary. We label the screens as numerical screens and report
the margin by which each passed, so the user can decide whether a
symbolic argument is worth making.

We begin with the exact operation. The Ma--Minda extremal function of
$\Sstar\varphi$ is
\begin{equation}\label{eq:extremal}
  f_\varphi(z) \;=\; z\exp\!\int_0^z \frac{\varphi(t) - 1}{t}\,dt
  \;=\; z\exp\Bigl(\sum_{k\ge1} \frac{B_k}{k} z^k\Bigr),
\end{equation}
and its coefficients follow from the recurrence for the exponential of a
series, \cref{alg:extremal}.

\begin{algorithm}[H]
\caption{\textsc{ExtremalCoefficients}$(\varphi, N)$}\label{alg:extremal}
\begin{algorithmic}[1]
\Require generator $\varphi$; order $1 \le N \le 24$
\Ensure exact coefficients $(a_2, \dots, a_{N+1})$ of $f_\varphi$ in \eqref{eq:extremal}
\State $(B_1, \dots, B_N) \gets$ \textsc{GeneratorSeries}$(\varphi, N)$
\State $e_0 \gets 1$
\For{$n = 1, \dots, N$}
  \State $e_n \gets \dfrac{1}{n}\displaystyle\sum_{k=1}^{n} B_k\, e_{n-k}$ \Comment{$f_\varphi/z = \sum e_n z^n$; exact rational arithmetic}
\EndFor
\State \Return $(e_1, \dots, e_N)$
\end{algorithmic}
\end{algorithm}

\begin{lstlisting}[caption={Exact extremal coefficients and the admissibility check.},label={lst:classes}]
from geometric_function_atlas import class_admissibility, class_extremal_coefficients

print(class_extremal_coefficients("exponential", order=6))
# (1, 3/4, 17/36, 19/72, 27/200, 8351/129600)
print(class_extremal_coefficients("sine", order=6))
# (1, 1/2, 1/9, -1/72, -4/225, -151/32400)

adm = class_admissibility("exponential")
print(adm.admissible, adm.exact_values)   # True {'phi0': '1', 'phi_prime0': '1'}
print({k: round(v, 5) for k, v in adm.margins.items()})
# {'re_min': 0.37158, 'symmetry_max_error': 0.0, 'starlike_wrt_1_min': 0.58537}
\end{lstlisting}

The admissibility check, \cref{alg:admiss}, decides $\varphi(0) = 1$ and
$\varphi'(0) > 0$ exactly, and evaluates the three region conditions of
\eqref{eq:generator} on four circles of radii $0.3, 0.6, 0.9, 0.99$ with
$120$ angles each. For the exponential class the smallest sampled value
of $\Re\varphi$ is $0.37158$, and the smallest sampled value of
$\Re\bigl(z\varphi'(z)/(\varphi(z)-1)\bigr)$, which is positive exactly
when the image is starlike with respect to $1$, is $0.58537$. Both
margins are comfortably positive.

\begin{algorithm}[H]
\caption{\textsc{ClassAdmissibility}$(\varphi)$}\label{alg:admiss}
\begin{algorithmic}[1]
\Require generator $\varphi$, from the catalogue or user-supplied
\Ensure verdict, exact values of $\varphi(0)$ and $\varphi'(0)$, and screen margins
\State decide exactly: $\varphi(0) = 1$; $\varphi'(0)$ real and positive
\State $G \gets \{\, r e^{2\pi i j/120} : r \in \{0.3, 0.6, 0.9, 0.99\},\ 0 \le j < 120 \,\}$
\State $m_1 \gets \min_{z\in G} \Re\varphi(z)$ \Comment{screen for $\Re\varphi > 0$}
\State $m_2 \gets \max_{z\in G,\ \Im z>0} \bigl|\varphi(\bar z) - \overline{\varphi(z)}\bigr|$ \Comment{screen for real coefficients}
\State $m_3 \gets \min_{z\in G} \Re\bigl(z\varphi'(z)/(\varphi(z)-1)\bigr)$ \Comment{screen for starlikeness w.r.t.\ $1$}
\State \Return admissible $\iff$ both exact checks pass, $m_1 > 0$, $m_2 < 10^{-9}$, $m_3 > -10^{-9}$; together with $(m_1, m_2, m_3)$
\end{algorithmic}
\end{algorithm}

The two remaining operations are screens by winding number. The
membership screen evaluates $zf'(z)/f(z)$ for a supplied polynomial on a
$16 \times 24$ polar grid with $|z| \le 0.95$ and tests each value
against the boundary $\varphi(0.999\,\partial\D)$ sampled at $720$
points. The containment screen samples the boundary of
$\varphi_{\mathrm{inner}}(0.99\,\D)$ at $180$ points and tests each point
against the sampled boundary of $\varphi_{\mathrm{outer}}(\D)$. By
transitivity of subordination, the second screen tests
$\Sstar{\varphi_{\mathrm{inner}}} \subseteq \Sstar{\varphi_{\mathrm{outer}}}$.
When it does not pass, it returns the boundary point that lies outside;
this point is a value of the extremal function of the inner class.

\begin{lstlisting}[caption={Membership and containment screens.},label={lst:screens}]
from geometric_function_atlas import class_member_screen, class_containment_screen

m = class_member_screen("exponential", [0.25, 0.10])     # f(z) = z + z^2/4 + z^3/10
print(m.member, m.fraction_inside, round(m.min_dist_to_boundary, 5))   # True 1.0 0.39627

c = class_containment_screen("exponential", "cardioid")
print(c.contained, c.fraction_inside, round(c.margin, 5))              # True 1.0 0.03824

c = class_containment_screen("sine", "sigmoid")
print(c.contained, c.fraction_inside, tuple(round(v, 5) for v in c.witness_w))
# False 0.0 (1.83603, 0.0)
\end{lstlisting}

The last call already points to \cref{sec:case}. The full sine domain is
not contained in the sigmoid domain, and the sampled point of
$\partial\varphi_{\sin}(0.99\,\D)$ that lies outside is
$1 + \sin(0.99) \approx 1.83603$, on the positive real axis. We label the
screens as numerical screens on purpose. They support exploration and
they locate the region where a symbolic argument is needed; the symbolic
argument itself is the subject of the radius certificates below.

\subsection{Tiered verification of membership}\label{sec:verify}

Let $f(z) = z + a_2 z^2 + \cdots + a_n z^n$ be a normalised polynomial
with real coefficients, and let $\mathcal P$ be one of the five supported
properties: starlikeness, convexity, univalence, and the Becker and
Nehari univalence criteria. Each of these except plain univalence has a
pointwise criterion $S_{\mathcal P}$ and a threshold. For starlikeness
$S(z) = \Re\bigl(zf'(z)/f(z)\bigr) > 0$. For convexity
$S(z) = \Re\bigl(1 + zf''(z)/f'(z)\bigr) > 0$. For the Becker-type
criterion $(1-|z|^2)\,|f''(z)/f'(z)| \le 1$, which implies Becker's
univalence condition~\cite{Becker} because $|z| < 1$. For the Nehari
criterion~\cite{Nehari} $(1-|z|^2)^2\,|S_f(z)| \le 2$, where $S_f$ is the
Schwarzian derivative. Starlikeness and convexity also have the classical
coefficient conditions
\begin{equation}\label{eq:c01}
  \Sigma_1 := \sum_{k\ge2} k\,|a_k| \;\le\; 1 \;\Longrightarrow\; f \in \mathcal S^*,
  \qquad
  \Sigma_2 := \sum_{k\ge2} k^2\,|a_k| \;\le\; 1 \;\Longrightarrow\; f \in \mathcal K,
\end{equation}
the first due to Schild and Goodman~\cite{Schild,Goodman}, the second to
Alexander~\cite{Alexander}. For a finite polynomial each is a complete
proof of the property. The condition that governs a symbolic proof
depends on the property: $\Sigma_1$ governs starlikeness, and hence
univalence; $\Sigma_2$ governs convexity; the Becker and Nehari criteria
have no such condition, so the symbolic tier makes no proof claim for
them. Our verifier, \cref{alg:verify}, answers at the tier the caller
requests.

\begin{algorithm}[H]
\caption{\textsc{VerifyFunction}$(a, \mathcal P, \tau)$}\label{alg:verify}
\begin{algorithmic}[1]
\Require coefficients $a = (a_2,\dots,a_n)$ as binary64 reals, read as the exact dyadic rationals they denote; property $\mathcal P$; tier $\tau \in \{\textsc{screen}, \textsc{symbolic}, \textsc{rigorous}\}$; grid $40 \times 48$ on $0.05 \le |z| \le 0.98$
\Ensure outcome, evidence kind, verification report, and for the rigorous tier a witness point with an enclosure
\State $\Sigma_{\mathcal P} \gets \Sigma_1$ for starlikeness and univalence, $\Sigma_2$ for convexity, undefined for the Becker and Nehari criteria \Comment{exact rational arithmetic; both sums are reported}
\If{$\tau = \textsc{screen}$}
  \State evaluate $S_{\mathcal P}$ in floating point on the grid; $m \gets$ worst margin against the threshold; $z_0 \gets$ its location
  \State \Return \out{passes screen} if $m \ge 0$ else \out{fails screen}; evidence \emph{numerical screen}; $(m, z_0)$
\EndIf
\If{$\tau = \textsc{symbolic}$}
  \If{$\Sigma_{\mathcal P}$ is undefined} \Return \out{criterion not symbolic}; evidence \emph{inconclusive}
  \ElsIf{$\Sigma_{\mathcal P} \le 1$ and $f$ is a finite polynomial} \Return \out{proven}; evidence \emph{exact proof} \Comment{by \eqref{eq:c01}}
  \ElsIf{$\Sigma_{\mathcal P} \le 1$} \Return \out{inconclusive truncation}
  \Else{} \Return \out{sufficient condition fails}, naming the condition; evidence \emph{inconclusive}
  \EndIf
\EndIf
\State grid scan as in the screen tier; $z_0 \gets$ worst point
\State $[\ell, u] \gets$ enclosure of $S_{\mathcal P}(z_0)$ by outward-rounded interval arithmetic \Comment{\cref{alg:witness}}
\If{the enclosure lies on the violating side of the threshold} \Return \out{certified violation}; evidence \emph{certified enclosure}; $(z_0, [\ell,u])$
\ElsIf{$\Sigma_{\mathcal P}$ is defined, $\Sigma_{\mathcal P} \le 1$ and $f$ is a finite polynomial} \Return \out{proven}; evidence \emph{exact proof}
\Else{} \Return \out{no certified violation on grid}; evidence \emph{numerical screen}
\EndIf
\end{algorithmic}
\end{algorithm}

Three points about \cref{alg:verify} deserve comment. First, we read the
coefficients as the exact dyadic rationals that their binary64
representations denote, so the exact sums are computed on the numbers
actually supplied. The input $0.1$ is the rational
$3602879701896397/36028797018963968$, and the envelope records it as
such. Second, the symbolic tier proves membership only through the
condition that governs the property, and both sums are reported in the
verification report with a truthful status: a sum above $1$ is marked as
failed. When the governing condition does not hold, the outcome names
it; this is a statement about the condition and not about $f$. Third,
the rigorous tier evaluates the enclosure at the worst screened point
before anything else. A certified violation stands whatever the sums say,
and only when no violation is certified does the governing condition
prove the property for a finite polynomial.

\begin{lstlisting}[caption={The three tiers on the polynomials $z + z^2/4$, $z + z^2/10$, $z + z^2$ and $z + 3z^2/10$.},label={lst:verify}]
from geometric_function_atlas import verify_function

s = verify_function([0.25], property="starlike", max_cost="screen")
print(s.outcome, s.details["grid_points"], round(s.min_margin, 4))
# passes_screen 1920 0.6755

p = verify_function([0.1], property="starlike", max_cost="symbolic")
print(p.outcome, p.evidence_kind, p.details["c01_sum"])
# proven exact_proof 3602879701896397/18014398509481984

r = verify_function([1.0], property="starlike", max_cost="rigorous")
print(r.outcome, tuple(round(v, 5) for v in r.witness_point))
# certified_violation (-0.98, 0.0)
print(round(r.details["interval_lower"], 5), round(r.details["interval_upper"], 5))
# -48.0 -48.0

c = verify_function([0.3], property="convex", max_cost="symbolic")
print(c.outcome)                  # alexander_fails_sufficient_condition
c = verify_function([0.3], property="convex", max_cost="rigorous")
print(c.outcome, tuple(round(v, 5) for v in c.witness_point))
# certified_violation (-0.98, 0.0)
\end{lstlisting}

The third example is worth a closer look. For $f(z) = z + z^2$ we have
$zf'(z)/f(z) = (1 + 2z)/(1 + z)$, which at $z = -0.98$ equals
$(1 - 1.96)/0.02 = -48$. The grid finds this point, the enclosure
brackets $-48$ to within $4\cdot10^{-14}$ (we print its endpoints above
rounded to five decimals), and its upper endpoint is negative, so the
violation is certified. The exact sum $\Sigma_1 = 2$ exceeds $1$, so the
coefficient condition gives no proof, and the verdict rests on the
enclosure alone. The fourth example shows why the governing condition
must depend on the property. For $f(z) = z + 0.3 z^2$ the starlikeness
sum is $\Sigma_1 = 0.6 \le 1$, so $f$ is starlike, while the convexity
sum is $\Sigma_2 = 1.2 > 1$; and indeed $1 + zf''(z)/f'(z)$ equals
$-44/103 < 0$ at $z = -0.98$, so $f$ is far from convex. The symbolic
tier reports that the convexity condition does not apply, and the
rigorous tier certifies the violation.

\subsection{Counterexample witnesses}\label{sec:cex}

A negative membership claim is settled by one point $z_0 \in \D$ where the
pointwise criterion fails. We offer the two natural operations on such
witnesses. One certifies a point supplied by the caller; the other
locates a candidate by a $120 \times 180$ grid search and then certifies
it. Both use \cref{alg:witness}, which is also the interval step of
\cref{alg:verify}, and both support starlikeness, convexity, and the
Becker and Nehari criteria.

\begin{algorithm}[H]
\caption{\textsc{CertifyWitness}$(a, \mathcal P, z_0)$}\label{alg:witness}
\begin{algorithmic}[1]
\Require coefficients $a$; property $\mathcal P \in \{$starlike, convex, Becker, Nehari$\}$; point $z_0 \in \D$ with binary64 coordinates
\Ensure enclosure $[\ell, u]$ of the criterion at $z_0$, threshold $t$, and \out{certified} $\in \{$true, false$\}$
\State $Z \gets$ thin complex interval at $z_0$; $F, F', F'', F''' \gets$ Horner evaluations of $f$ and its derivatives at $Z$ in interval arithmetic, the integer factors $k$, $k(k-1)$, $k(k-1)(k-2)$ multiplied inside the intervals
\If{the relevant denominator interval ($F$ for starlikeness, $F'$ otherwise) contains $0$} \textbf{fail} with the state \out{unresolved} \EndIf
\State $Q \gets Z F'/F$, or $1 + Z F''/F'$, or $(1-|Z|^2)\,|F''/F'|$, or $(1-|Z|^2)^2\,|F'''/F' - \tfrac32 (F''/F')^2|$ according to $\mathcal P$
\State $[\ell, u] \gets$ endpoints of $\Re Q$ (starlike, convex) or of $Q$ (Becker, Nehari); $t \gets 0$, $0$, $1$, or $2$ respectively
\State \Return \out{certified} $\iff u \le t$ (starlike, convex) or $\ell > t$ (Becker, Nehari)
\end{algorithmic}
\end{algorithm}

\begin{lstlisting}[caption={Certifying a supplied witness and locating one by search.},label={lst:cex}]
from geometric_function_atlas import find_counterexample, verify_counterexample

w = verify_counterexample([1.0], point=(-0.75, 0.0), property="starlike")
print(w.certified, w.interval_lower, w.interval_upper, w.direction)
# True -2.0 -2.0 disproves

s = find_counterexample([1.0], property="starlike")
print(s.certified, tuple(round(v, 5) for v in s.point), s.grid_points)
# True (-0.99, 0.0) 21600
print(round(s.interval_lower, 5), round(s.interval_upper, 5))
# -98.0 -98.0

k = verify_counterexample([0.6], point=(-0.813, 0.0), property="convex")
print(k.certified, round(k.interval_upper, 5), k.direction)
# True -38.98361 disproves
\end{lstlisting}

At $z_0 = -3/4$ the value $(1 + 2z_0)/(1 + z_0) = -2$ is a dyadic
rational. Because the inputs are dyadic, the interval arithmetic
reproduces it exactly, and the enclosure is the single point $[-2, -2]$.
The search finds the most violating grid point instead, $z = -0.99$,
where the value is $-98$. The last call certifies that $z + 0.6 z^2$ is
not convex: at $z = -0.813$ the derivative $f'(z) = 1 + 1.2z$ is close to
zero, and $1 + zf''(z)/f'(z)$ is a large negative number. A certified
witness is the strongest evidence against a positive claim, since it is a
machine-checkable disproof. The envelope carries the point, the enclosure
and the threshold, so the check can be repeated later.

\subsection{Directed inclusion radii}\label{sec:radii}

Given two Ma--Minda generators $\varphi_1$ and $\varphi_2$, the directed
radius problem asks for the largest $r \in (0,1]$ such that every
$f \in \Sstar{\varphi_1}$, dilated to the disk of radius $r$, lies in
$\Sstar{\varphi_2}$. Write $f_r(z) = r^{-1} f(rz)$ and $w = zf'/f$. Then
$z f_r'(z)/f_r(z) = w(rz)$, and $f_r \in \Sstar{\varphi_2}$ exactly when
$w(r\D) \subseteq \varphi_2(\D)$. Since $w \sub \varphi_1$ gives
$w(r\D) \subseteq \varphi_1(r\D)$, with equality for the extremal
function \eqref{eq:extremal} of the source class, the radius is
\begin{equation}\label{eq:radius}
  r^\star(\varphi_1 \to \varphi_2)
  \;=\;
  \sup\bigl\{\, r \in (0,1] \;:\; \varphi_1(r\D) \subseteq \varphi_2(\D) \,\bigr\}.
\end{equation}
Sharpness is witnessed by $f_{\varphi_1}$ together with the point of
$\partial\D$ where $\varphi_1(r^\star z)$ touches $\partial\varphi_2(\D)$.
Direction matters in \eqref{eq:radius}. We record, for instance,
$r^\star(\sin \to \mathrm{sigmoid}) = \arcsin\bigl((e-1)/(e+1)\bigr)$ and
$r^\star(\mathrm{sigmoid} \to \sin) = 1$.

We treat each ordered pair as a row of an immutable snapshot of $702$
directed radii, and we give each row one of five statuses that record the
evidence behind its value. At the current snapshot $323$ rows have an
exactly solved contact equation on the axis, with the angular maximum
confirmed on an $8192$-point grid. A further $140$ rows carry a closed
form that agrees with a high-precision value to about one hundred digits.
Another $142$ are trivial containments with $r^\star = 1$. A further $85$
carry a sixty-digit value with no closed form assigned yet; they form the
reservoir of candidate closed forms discussed in \cref{sec:future}. The
remaining $12$ are held outside the reconcilable set for audit. Eight
rows, listed in \cref{tab:lanes}, also carry a reviewed certificate. A
certificate is a recorded chain of steps that reduces the containment
\eqref{eq:radius} to a threshold equation on the axis, together with the
assumptions, the inverse branch of $\varphi_2$, and the attainment
statement. We replay such a chain with \cref{alg:replay}.

\begin{algorithm}[H]
\caption{\textsc{ReplayRadiusCertificate}$(\mathrm{record}, c, p)$}\label{alg:replay}
\begin{algorithmic}[1]
\Require radius record for $(\varphi_1, \varphi_2)$; optional candidate expression $c$; precision $16 \le p \le 200$ digits
\Ensure status $\in \{$\out{proven}, \out{unresolved}, \out{candidate mismatch}, \out{invalid input}, \out{unsupported}, \out{corrupt artifact}$\}$ and the list of steps with verdicts
\If{the record carries no certificate} \Return \out{unsupported} \Comment{a snapshot row without a reviewed chain}\EndIf
\State check that the direction is among the eight reviewed lanes, that the source commit, crosswalk commit, fixture digest and lane identity match the bundled artifact, and that the record equals the trusted snapshot row; otherwise \Return \out{corrupt artifact}
\State parse $c$ (default: the stored exact radius) with a closed grammar admitting integers, $e$, $\pi$, and the functions $\sin,\cos,\tan,\sinh,\cosh,\tanh,\exp,\log,\sqrt,\arcsin,\operatorname{arsinh},\operatorname{artanh},\operatorname{arcosh}$; nesting depth at most $32$
\If{$\operatorname{simplify}(c - c_{\mathrm{stored}}) \ne 0$} \Return \out{candidate mismatch} \EndIf
\ForAll{steps $s$ of the lane's chain}
  \State \textbf{identity step:} verified $\iff \operatorname{simplify}(\mathrm{lhs}_s - \mathrm{rhs}_s) = 0$ in SymPy
  \State \textbf{threshold step:} verified $\iff |\,\mathrm{N}(\mathrm{lhs}_s - \mathrm{rhs}_s,\ p)\,| < 10^{-\min(30,\, p/2)}$
  \State \textbf{classical step:} a standard identity or inequality (the logarithmic form of $\operatorname{artanh}$, a triangle inequality, or a series majorant) recorded with its statement
\EndFor
\State \Return \out{proven} if every step verified, else \out{unresolved}
\end{algorithmic}
\end{algorithm}

We expose the pieces of this machinery through eight radius operations.
Two read the snapshot, one row or many. Four run \cref{alg:replay} on a
record or on a named pair: the plain replay, a replay by name, a
recomputation, and the attainment check. One identifies the snapshot rows
whose exact value equals a supplied expression symbolically, or whose
float value lies within a tolerance of a supplied decimal. One bundles a
replay with the record's evidence and claim labels for audit.

\begin{lstlisting}[caption={Reading, replaying, and identifying directed radii.},label={lst:radii}]
from geometric_function_atlas import (
    identify_radius, list_radii, radius, replay_radius_certificate,
)

print(len(list_radii()), len(list_radii(status="touch_proven_exact")))   # 702 323

rec = radius("sine", "sigmoid")
print(rec.value_exact, rec.status.value, rec.touch_angle, rec.mode)
# asin((E-1)/(E+1)) touch_proven_exact 0.0 axis
print(round(rec.value_float, 5))        # 0.48038

rep = replay_radius_certificate(rec)    # default precision 50 digits
print(rep.status, rep.certified, len(rep.steps))          # proven True 5

bad = replay_radius_certificate(rec, candidate="asinh((E-1)/(E+1))")
print(bad.status)                                          # candidate_mismatch

print([r.direction for r in identify_radius(0.48038, tolerance=1e-5)])   # ['sine->sigmoid']
print(len(identify_radius(0.5)))               # 14
print(radius("sigmoid", "sine").value_exact)   # 1
\end{lstlisting}

We accept a candidate that is a different closed form of the same
number, because we decide equality by symbolic simplification. Identifying
by the expression $\arcsin(\tanh(1/2))$ returns the same lane, because
$\tanh(1/2) = (e-1)/(e+1)$. Identification by decimal value uses a
tolerance supplied by the caller, and the value $1/2$ is shared by
fourteen directed pairs of the snapshot. We report a candidate that is a
different number, such as the published sufficient radius
$\operatorname{arsinh}\bigl((e-1)/(e+1)\bigr)$ tried above, as a
mismatch. \Cref{sec:case} follows the sine-to-sigmoid lane through the
chain in detail.

\subsection{Certificates, expansions and bounds}\label{sec:artifacts}

Our artifact snapshot also carries the coefficient side of our registry.
One operation lists the $306$ coefficient certificates. Another returns
one certificate with its bound, its slack, the engine that produced it,
the classical lemmas it uses, the number of leaves of its
interval-arithmetic partition, and its stored extremal Schur parameters.
Another looks up the baked bound for a class and functional. Another
returns the Schur-parameter expansion of the low-order coefficients of a
class. Another lists the $12$ certified enclosures and $97$ numerical
conjectures of the open-problem table. Another returns the $227$-row
table that compares each certificate with the extracted literature. The
operation that turns this data into evidence is the certificate replay,
\cref{alg:verifycert}. It re-executes the stored extremal of a
certificate through the exact Schur machinery and confirms that the
functional attains the stored bound there.

\begin{algorithm}[H]
\caption{\textsc{VerifyCertificate}$(\mathrm{name})$}\label{alg:verifycert}
\begin{algorithmic}[1]
\Require certificate name, for instance the starlike Fekete--Szeg\H{o} certificate at $\mu = 1$
\Ensure matched flag, exact functional value at the extremal, and a verification report
\State look up the certificate in the checksummed snapshot; require declared status \textsc{proved}
\State $\gamma \gets$ stored extremal Schur parameters; $(a_2, a_3, \dots) \gets$ exact coefficients of the member of $\Sstar\varphi$ with Schur parameters $\gamma$
\State $v \gets$ exact value of the functional at $(a_2, a_3, \dots)$
\State record checks: artifact present; upper bound certified in the source; $v$ equals the declared candidate; $v$ consistent with the declared bound; sharpness flag consistent
\State \Return matched $\iff$ every required check passed, together with $v$ and the report
\end{algorithmic}
\end{algorithm}

\begin{lstlisting}[caption={Snapshot identity and certificate replay.},label={lst:artifacts}]
from geometric_function_atlas import (
    coefficient_bound, expansion, get_proof, list_proofs,
    snapshot_info, snapshot_verify, verify_certificate,
)

info = snapshot_info()
print(info["artifact_version"], len(info["files"]))          # 2026.08.11 8
print(snapshot_verify()["files_verified"], snapshot_verify()["checks"]["success"])   # 8 True

print(list_proofs()["count"])                                 # 306
p = get_proof("starlike__fekete_szego_mu1")
print(p["status"], p["bound"], p["n_leaves"], p["sharp"]["extremal_omega"])
# PROVED 1 1556938 z^2

v = verify_certificate("starlike__fekete_szego_mu1")
print(v["matched"], v["functional_value_exact"], v["sharpness_proven"])   # True 1 True

print(coefficient_bound("starlike", "fekete_szego_mu1")["bound"])   # 1
print(expansion("exponential")["coeffs"][1]["latex"])
# \frac{3 \gamma_{0}^{2}}{4} - \frac{\gamma_{0} \gamma_{1} \bar{\gamma}_{0}}{2} + \frac{\gamma_{1}}{2}
\end{lstlisting}

Our certificate for $|a_3 - a_2^2|$ over the classical starlike class
records the bound $1$, a proof by an interval-arithmetic partition of
$1{,}556{,}938$ leaves over the Schur-parameter body, and the extremal
Schwarz function $\omega(z) = z^2$. The replay recomputes the functional
at the stored parameters and finds exactly $1$. The expansion shown is
the coefficient $a_3$ of a member of $\Sstar{e^z}$ in terms of its Schur
parameters $\gamma_0, \gamma_1$; this is the object on which such
certificates operate.

We distribute the relational registry of papers, families and facts
separately, as a SQLite artifact that we keep out of the wheel on
purpose. A user who obtains a database and its manifest verifies and
installs it with two operations, and queries it through a snapshot object
whose methods mirror the pages of our public website: statistics, search,
families, papers, facts, evidence, counterexamples, aliases and the
property hierarchy.

\subsection{Figures}\label{sec:plot}

\begin{figure}[H]
\centering
\scalebox{1.15}{\input{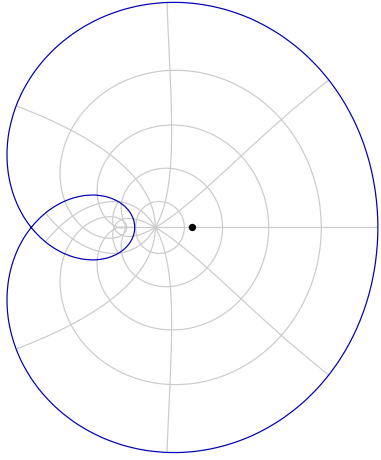}}
\caption{The domain figure for the sine generator, written by the package
as TikZ: the image of a polar grid in $\D$ under the Taylor polynomial of
$z(1 + \sin z)$ of order $12$, with the image of the unit circle in blue.}
\label{fig:domain}
\end{figure}

Our plotting module writes dependency-free SVG figures of four kinds: the
image of a polar grid under the function, the coefficient magnitudes, a
heat map of $\Re f$, and a phase portrait, for a catalogue generator or
for supplied coefficients. The plotted function is the Taylor polynomial
of $z\varphi(z)$ of order $12$. The figures are visual aids, and we label
them as such. We can also write the domain figure as PNG or as TikZ
source for direct use in a manuscript; \cref{fig:domain} is such a TikZ
figure, included here unchanged.

\begin{lstlisting}[caption={Writing figures.},label={lst:plot}]
from geometric_function_atlas import write_plot

out = write_plot("domain", "exp-domain.svg", generator="exponential")
print(out.approximation)                   # f(z) = z*phi(z), Taylor order 12
write_plot("domain", "sine-domain.tikz", generator="sine")
write_plot("phase",  "exp-phase.svg",  generator="exponential")
\end{lstlisting}

\section{What each operation establishes}\label{sec:guarantees}

The operations of \cref{sec:features} return labelled outcomes, and it is
natural to ask what each label proves. We record the answers as
propositions. Throughout, $a = (a_2, \dots, a_n)$ is a vector of binary64
reals read as the exact dyadic rationals they denote, and
$f(z) = z + \sum a_k z^k$.

\begin{proposition}[Exact computations]\label{prop:exact}
\textsc{GeneratorSeries}, \textsc{FeketeSzego} and
\textsc{ExtremalCoefficients} return exact elements of $\Q$ or of an
algebraic extension of $\Q$. In particular, for a catalogue generator
with $B_1 > 0$ real and $B_2$ real, the value returned by
\textsc{FeketeSzego}$(\varphi, \mu)$ equals the right-hand side of
\eqref{eq:fs-bound}, which is the sharp maximum of $|a_3 - \mu a_2^2|$
over $\Sstar\varphi$ under the Ma--Minda hypotheses on $\varphi$.
\end{proposition}

\begin{proposition}[Screen tier]\label{prop:screen}
If \textsc{VerifyFunction}$(a, \mathcal P, \textsc{screen})$ returns
\out{passes screen}, then the floating-point value of the criterion
$S_{\mathcal P}$ lies on the admissible side of its threshold at each of
the $1920$ grid points $r_j e^{i\theta_k}$ with
$r_j = 0.05 + 0.93\,j/39$ and $\theta_k = 2\pi k/48$. We assert nothing
about other points of $\D$, and the label \emph{numerical screen} records
this.
\end{proposition}

\begin{proposition}[Symbolic tier]\label{prop:symbolic}
If \textsc{VerifyFunction}$(a, \textsc{starlike}, \textsc{symbolic})$
returns \out{proven}, then $\Sigma_1 \le 1$ holds in exact arithmetic,
and therefore $\Re\bigl(zf'(z)/f(z)\bigr) > 0$ on $\D$, so
$f \in \mathcal S^*$; the same outcome for \textsc{univalent} gives
univalence, since starlike functions are univalent. If
\textsc{VerifyFunction}$(a, \textsc{convex}, \textsc{symbolic})$ returns
\out{proven}, then $\Sigma_2 \le 1$ holds in exact arithmetic, and
$f \in \mathcal K$. For the Becker and Nehari criteria the symbolic tier
returns \out{criterion not symbolic} and makes no proof claim. If the
governing sum exceeds $1$, the outcome names the condition that did not
hold, and we make no statement about $f$.
\end{proposition}

\begin{proposition}[Rigorous tier and witnesses]\label{prop:rigorous}
If \textsc{VerifyFunction}$(a, \textsc{starlike}, \textsc{rigorous})$
returns \out{certified violation} with witness $z_0$ and enclosure
$[\ell, u]$, then $z_0 \in \D$ and $\Re\bigl(z_0 f'(z_0)/f(z_0)\bigr) \le u \le 0$,
so $f \notin \mathcal S^*$. The corresponding statement for
\textsc{convex} gives $f \notin \mathcal K$. The same conclusions hold
when \textsc{CertifyWitness} returns \out{certified} for a point supplied
by the caller or found by search. For the Becker-type and Nehari
criteria, a certified violation shows that the criterion fails at $z_0$.
Both criteria are sufficient conditions for univalence, so nothing
follows about the univalence of $f$; we label the direction of such a
witness \emph{violates criterion} and reserve the word \emph{disproves}
for starlikeness and convexity. An outcome of \out{proven} at the
rigorous tier has exactly the content of \cref{prop:symbolic}.
\end{proposition}

\begin{proposition}[Radius certificate replay]\label{prop:radius}
Let the record for $(\varphi_1, \varphi_2)$ be one of the eight reviewed
lanes, and suppose \textsc{ReplayRadiusCertificate} returns
\out{proven} for candidate $c$ at precision $p$. Then
\begin{enumerate}[leftmargin=*,itemsep=1pt]
  \item $c$ equals the stored exact radius symbolically, and the
        record's provenance digests match the bundled artifact;
  \item every identity step of the chain holds, discharged by symbolic
        simplification, and the threshold equation holds to at least
        $\min(30, p/2)$ digits;
  \item the chain's classical step is a standard identity or inequality
        whose statement is recorded in the certificate.
\end{enumerate}
Under the assumptions recorded in the certificate, namely the
normalisation of the two classes, the Schwarz-lemma dilation, the
principal branch of $\varphi_2^{-1}$, and the closed-disk convention for
$r^\star$, the chain is the machine-checkable content of the proof that
$\varphi_1(r\D) \subseteq \varphi_2(\D)$ for $r < r^\star$ with contact at
$r^\star$, as we give it in our companion paper.
\end{proposition}

\begin{proposition}[Coefficient certificate replay]\label{prop:cert}
If \textsc{VerifyCertificate}$(\mathrm{name})$ returns matched, then the
functional named by the certificate, evaluated exactly at the member of
$\Sstar\varphi$ with the stored extremal Schur parameters, equals the
stored candidate value, and that value agrees with the bound whose
upper-bound certification the source artifact records.
\end{proposition}

The propositions say exactly what each operation establishes, and our
labels make the same distinctions visible to the user. The rigorous tier
gives a pointwise disproof, and we establish membership through
\cref{prop:symbolic}. The radius replay re-executes the recorded proof
chain of a lane; its classical steps are standard identities and
inequalities, and we give its analytic proof in our companion paper. We
likewise record the upper-bound certification of a coefficient
certificate in the source artifact and prove it there.

\section{A worked case: the sine-to-sigmoid radius}\label{sec:case}

To make the preceding sections concrete, we follow one lane of the radius
atlas from the exploratory screen through the exact reduction to the
replayed certificate, and we place the resulting value beside the
published sufficient radius that it improves. The lane is
$\Sstar{\varphi_{\sin}} \to \Sstar{\varphi_{\mathrm{sig}}}$ with
\[
  \varphi_{\sin}(z) \;=\; 1 + \sin z,
  \qquad
  \varphi_{\mathrm{sig}}(z) \;=\; \frac{2}{1 + e^{-z}},
\]
the sine class of Cho, Kumar, Kumar and Ravichandran~\cite{ChoSine} and
the modified-sigmoid class of Goel and Kumar~\cite{GoelKumar}.

\paragraph{Screening.}
The containment screen of \cref{sec:classes} reports that the full sine
domain does not fit inside the sigmoid domain. None of the $180$ sampled
points of $\partial\varphi_{\sin}(0.99\,\D)$ lies inside
$\varphi_{\mathrm{sig}}(\D)$, and the returned witness
$1 + \sin(0.99) \approx 1.83603$ lies on the positive real axis, beyond the
right endpoint $\varphi_{\mathrm{sig}}(1) = 2e/(e+1) \approx 1.46212$ of the
sigmoid domain. So the radius is strictly less than one, and the screen
suggests that the first contact occurs on the axis.

\paragraph{Reduction.}
The principal inverse of the sigmoid is $\varphi_{\mathrm{sig}}^{-1}(w) =
\log\bigl(w/(2-w)\bigr)$, so the composition
$\psi = \varphi_{\mathrm{sig}}^{-1}\circ\varphi_{\sin}$ is
\begin{equation}\label{eq:psi}
  \psi(z) \;=\; \log\frac{1 + \sin z}{1 - \sin z} \;=\; 2\operatorname{artanh}(\sin z).
\end{equation}
By \eqref{eq:radius}, the containment $\varphi_{\sin}(r\D) \subseteq
\varphi_{\mathrm{sig}}(\D)$ is the same as $|\psi(z)| \le 1$ for
$|z| \le r$. Write $z = x + iy$. The identity
$|\cos z|^2 = \cos^2 x + \sinh^2 y$ shows that $|\cos z| \ge |\cos x|$,
with difference $\sinh^2 y \ge 0$. Together with
$\tfrac{d}{dz}\operatorname{artanh}(\sin z) = \sec z$, this bounds the
angular variation of $\psi$ on the circle $|z| = r$ by its value on the
axis. On the axis $\psi(r) = 2\operatorname{artanh}(\sin r)$ increases
with $r$, so the radius solves $2\operatorname{artanh}(\sin r^\star) = 1$,
that is $\sin r^\star = \tanh\tfrac12 = (e-1)/(e+1)$, and
\begin{equation}\label{eq:sinsig}
  r^\star_{\sin\to\mathrm{sig}}
  \;=\;
  \arcsin\!\left(\frac{e-1}{e+1}\right)
  \;\approx\; 0.48038 .
\end{equation}
Attainment comes from the extremal function \eqref{eq:extremal} of the
sine class, whose Schwarz witness is $\omega(z) = z$ and whose image
touches $\partial\varphi_{\mathrm{sig}}(\D)$ at $z = r^\star$.

\paragraph{Replay.}
These are exactly the steps we recorded in the lane's certificate, and
\cref{alg:replay} re-executes them:

\begin{lstlisting}[caption={Replaying the sine-to-sigmoid certificate.},label={lst:case-replay}]
from geometric_function_atlas import radius, replay_radius_certificate

rec = radius("sine", "sigmoid")
print(rec.certificate.exact_candidate)     # asin((E-1)/(E+1))
print(rec.certificate.inverse_branch_and_domain)
# psi(z)=2*atanh(sin z), principal atanh branch on the certified disk; ...
print(rec.certificate.contact_and_attainment)
# Positive real axis z=r; 2*atanh(sin r)=1. Equality is attained by the
# Ma-Minda extremal Schwarz witness omega(z)=z and the standard dilation argument.

rep = replay_radius_certificate(rec, dps=100)
for step in rep.steps:
    print(step.verified, step.name)
# True psi composition reduces to the logarithmic sine form
# True d/dz atanh(sin z) = sec z
# True |cos(x+iy)|^2 = cos^2 x + sinh^2 y
# True angular-bound remainder is sinh(b)^2 >= 0
# True 2*atanh(sin(r*)) = 1
print(rep.status, rep.certified)           # proven True
\end{lstlisting}

The second, third and fourth steps are discharged by symbolic
simplification, and the fifth is checked to one hundred digits. The
first records the classical identity
$\log\bigl((1+s)/(1-s)\bigr) = 2\operatorname{artanh} s$ that collapses
the composition \eqref{eq:psi} to its logarithmic form; the hand-written
check of \cref{sec:perf} discharges this identity in SymPy as well. By
\cref{prop:radius}, the replay is the machine-checkable content of the
proof of \eqref{eq:sinsig}. The other seven lanes have the same shape:
one classical inequality recorded with its statement, a monotonicity
statement discharged by differentiation, and a threshold equation.

\paragraph{Reconciliation with the literature.}
The literature status of the record is \emph{candidate improvement}. The
status has a precise meaning in the reconciliation engine of our research
repository. The engine joins each certified radius against the radius
claims we extracted from the literature corpus, matching by direction, by
generator and by numerical value. We assign this status when a published
sufficient radius exists for the same directed pair and our certified
sharp value exceeds it. For this lane the extracted claim is
Theorem~2.9(ii) of Goel and Kumar~\cite{GoelKumarRadius}, which gives the
sufficient radius $\operatorname{arsinh}\bigl((e-1)/(e+1)\bigr) \approx
0.44707$ through the enclosure $|\sin z| \le \sinh r$. Our sharp value
\eqref{eq:sinsig} exceeds it by a factor of $1.0745$, a relative
improvement of $7.45\%$. The replay above rejects the published value as
a candidate for the sharp radius, reporting a candidate mismatch, and
accepts any closed form symbolically equal to \eqref{eq:sinsig}. The same
paper states five further sigmoid radii as sharp, for the cardioid,
rational, exponential, lemniscate and crescent source classes, and our
engine reproduces all five exactly. Those rows carry the status
\emph{known} and serve as positive controls for the one row that improves
on its published counterpart.

Our reconciliation vocabulary keeps bibliographic relationships apart
from computational evidence on purpose. The certificate records the
radius as a machine-proven exact inclusion radius. The sharpness of
\eqref{eq:sinsig} rests on the replayed chain and on the proof in our
companion paper. The literature status records, separately, how the value
relates to the extracted corpus. For the same reason the novelty flag of
every envelope is false: our package reports computations and their
bibliographic context, and we make claims of novelty in our mathematical
papers. The status \emph{no extracted claim} carried by six of the
reviewed lanes is likewise a statement about the corpus: it contains no
radius claim for the pair.

\paragraph{The eight reviewed lanes.}
In \cref{tab:lanes} we list the eight lanes that carry replayable
certificates, with their exact values, their snapshot status, their
literature status, the number of recorded steps, and the replay time we
measured in \cref{sec:perf}. Two lanes, $\exp \to$ order $\tfrac12$ and
starlike $\to$ order $\tfrac34$, reproduce classical results and we label
them so. For the remaining six, the extracted corpus held no literature
claim to compare against at the snapshot date.

\begin{table}[H]
\centering
\footnotesize
\setlength{\tabcolsep}{4.5pt}
\begin{tabular}{lllccrr}
\toprule
Lane $\varphi_1 \to \varphi_2$ & $r^\star$ (exact) & $r^\star$ (decimal) & Status & Lit. & Steps & Replay (ms) \\
\midrule
sine $\to$ sigmoid                 & $\arcsin\frac{e-1}{e+1}$ & $0.48038$ & T & C & 5 & $29.6$ \\
sine $\to$ tanh                    & $\arcsin(\tanh 1)$       & $0.86577$ & F & N & 5 & $29.6$ \\
crescent $\to$ lemniscate          & $\sqrt2/4$               & $0.35355$ & T & N & 4 & $0.12$ \\
starlike $\to$ lemniscate          & $3 - 2\sqrt2$            & $0.17157$ & T & N & 4 & $2.71$ \\
order $\tfrac12 \to$ crescent      & $2 - \sqrt2$             & $0.58579$ & T & N & 6 & $3.82$ \\
exponential $\to$ order $\tfrac12$ & $\log 2$                 & $0.69315$ & T & K & 4 & $0.40$ \\
exponential $\to$ lemniscate       & $\tfrac12\log 2$         & $0.34657$ & T & N & 5 & $0.14$ \\
starlike $\to$ order $\tfrac34$    & $1/7$                    & $0.14286$ & T & K & 5 & $15.4$ \\
\bottomrule
\end{tabular}
\caption{The eight reviewed lanes of the radius snapshot
\texttt{gft-radius-snapshot:2026.08.09}. Decimals are rounded to five
places. Status: T, contact equation solved exactly on the axis; F, closed
form confirmed to about one hundred digits. Literature: C, improves an
extracted published bound; K, known general result; N, no extracted
claim. Steps is the number of recorded chain steps; the replay time is
the warm median of \cref{tab:perf-radius}. All eight replays return
\out{proven}.}
\label{tab:lanes}
\end{table}

\section{Measured performance and a hand-written comparison}\label{sec:perf}

We measured all timings in this section on a laptop with an Apple M3 Pro
processor (eleven cores, five of them performance cores), $18$~GB of
memory, macOS 26.6.2, CPython 3.12.9, SymPy 1.14.0, mpmath 1.3.0 and
NumPy 2.5.3, running \pkg{} \pkgversion. We ran each configuration in a
fresh interpreter, so peak resident memory belongs to that configuration
alone. We report the first call separately, because it includes loading
the bundled artifacts and verifying their digests, and then the median of
seven further calls. Importing our package took between $137$ and
$151$~ms in every process, and peak resident memory was between $66$ and
$70$~MB in every configuration; both are dominated by SymPy. Our
benchmark script and its JSON output accompany the manuscript.

\begin{table}[H]
\centering
\small
\begin{tabular}{lrrrr}
\toprule
Configuration & Precision (digits) & First call (ms) & Warm median (ms) & Peak RSS (MB) \\
\midrule
sine $\to$ sigmoid & $16$  & $82.2$ & $28.5$ & $70.2$ \\
sine $\to$ sigmoid & $50$  & $86.6$ & $29.6$ & $70.1$ \\
sine $\to$ sigmoid & $100$ & $86.5$ & $29.9$ & $70.0$ \\
sine $\to$ sigmoid & $200$ & $82.0$ & $28.4$ & $70.2$ \\
\midrule
sine $\to$ tanh                    & $50$ & $86.2$ & $29.6$ & $70.1$ \\
crescent $\to$ lemniscate          & $50$ & $10.2$ & $0.12$ & $67.9$ \\
starlike $\to$ lemniscate          & $50$ & $52.9$ & $2.71$ & $69.5$ \\
order $\tfrac12 \to$ crescent      & $50$ & $64.1$ & $3.82$ & $69.8$ \\
exponential $\to$ order $\tfrac12$ & $50$ & $36.6$ & $0.40$ & $69.3$ \\
exponential $\to$ lemniscate       & $50$ & $6.1$  & $0.14$ & $67.9$ \\
starlike $\to$ order $\tfrac34$    & $50$ & $74.1$ & $15.4$ & $69.8$ \\
\bottomrule
\end{tabular}
\caption{Radius certificate replay (\cref{alg:replay}). The first block
varies the precision of the threshold check for one lane; the second
replays each reviewed lane at the default precision.}
\label{tab:perf-radius}
\end{table}

The first block of \cref{tab:perf-radius} shows that the replay time is
independent of the requested precision, across the whole admissible range
from $16$ to $200$ digits. This is as expected. All but one step of a
chain are symbolic identities discharged by SymPy's simplifier, and the
single numerical step evaluates one elementary expression, whose cost at
two hundred digits is negligible next to the symbolic work. The second
block shows that the cost depends on the identities in the chain; the
difficulty of the lane as a theorem plays no part. The two lanes whose
chains involve $\operatorname{artanh}(\sin z)$ take about $30$~ms, the
lane with a trigonometric modulus identity about $15$~ms, and the lanes
with rational or exponential identities well under $5$~ms.

\begin{table}[H]
\centering
\small
\begin{tabular}{llrr}
\toprule
Operation & Argument & First call (ms) & Warm median (ms) \\
\midrule
Fekete--Szeg\H{o} constant        & exponential, $\mu = 0$        & $7.8$  & $5.3$ \\
Generator series                  & exponential, order $4$        & $11.2$ & $7.4$ \\
Generator series                  & exponential, order $12$       & $21.5$ & $13.8$ \\
Extremal coefficients             & exponential, order $8$        & $16.7$ & $11.3$ \\
Extremal coefficients             & exponential, order $24$       & $40.7$ & $25.9$ \\
Admissibility check               & exponential ($480$ points)    & $3.5$  & $0.57$ \\
Membership screen                 & exponential, $[0.25, 0.1]$    & $30.5$ & $28.9$ \\
Containment screen                & exponential $\to$ cardioid    & $16.7$ & $14.7$ \\
Verifier, screen tier             & $1920$ points                 & $1.60$ & $1.47$ \\
Verifier, symbolic tier           &                               & $0.27$ & $0.10$ \\
Verifier, rigorous tier           &                               & $1.71$ & $1.54$ \\
Witness certification             & $z + z^2$ at $-3/4$           & $0.13$ & $0.06$ \\
Witness search                    & $z + z^2$, $21{,}600$ points  & $19.1$ & $18.5$ \\
Coefficient certificate replay    & starlike, $\mu = 1$           & $4.39$ & $0.08$ \\
Snapshot digest check             & eight files, SHA-256          & $6.99$ & $3.41$ \\
\bottomrule
\end{tabular}
\caption{Measured cost of the remaining operations of \cref{sec:features}
on the same machine. Every operation completes in well under one tenth of
a second; the screens are bounded by the number of floating-point
evaluations and the exact operations by SymPy's rational arithmetic.}
\label{tab:perf}
\end{table}

\paragraph{A hand-written comparison.}
A reader who knows what to check can verify the identities of a radius
chain directly in SymPy, and it is useful to see what that involves.
\Cref{lst:handrolled} verifies all five steps of the sine-to-sigmoid
chain by hand, including the first, which our chain records as a
classical identity and which SymPy discharges through the logarithmic
form of $\operatorname{artanh}$. The hand-written check agrees with our
replay on every step. In the same interpreter the two took warm medians
of $28.4$~ms and $28.1$~ms. Our replay therefore adds no measurable
overhead to the symbolic work.

\begin{lstlisting}[caption={Hand-written verification of the sine-to-sigmoid identities in SymPy.},label={lst:handrolled}]
import sympy as sp
z = sp.Symbol("z"); s = sp.Symbol("s", positive=True)
x, y, a, b, rho = sp.symbols("x y a b rho", real=True)

lhs = sp.expand_log(sp.log((1 + s)/(1 - s)), force=True)          # s = sin z
rhs = sp.expand_log((2*sp.atanh(s)).rewrite(sp.log), force=True)
psi_ok = sp.simplify(lhs - rhs) == 0                               # psi = 2 artanh(sin z)
sec_ok = sp.simplify(sp.diff(sp.atanh(sp.sin(z)), z) - sp.sec(z)) == 0
mod_ok = sp.simplify(sp.expand_complex(sp.Abs(sp.cos(x + sp.I*y))**2)
                     - (sp.cos(x)**2 + sp.sinh(y)**2)) == 0
rem    = (sp.cos(a)**2 + sp.sinh(b)**2 - sp.cos(rho)**2) - (sp.cos(a)**2 - sp.cos(rho)**2)
rem_ok = sp.simplify(rem - sp.sinh(b)**2) == 0
r_star = sp.asin((sp.E - 1)/(sp.E + 1))
thr_ok = abs(sp.N(2*sp.atanh(sp.sin(r_star)) - 1, 50)) < sp.Float(10)**-25
print(psi_ok, sec_ok, mod_ok, rem_ok, thr_ok)      # True True True True True
\end{lstlisting}

What the hand-written check lacks is everything around the identities.
Our replay adds the recorded assumptions and inverse branch, the
attainment statement, the pinning of the chain to a source commit and a
fixture digest, the rejection of a candidate that is a different number,
and the envelope that reports all of this in a fixed vocabulary. A
referee needs these components, and our replay supplies them in one call
at the same cost as the bare identities.

\section{Reproducibility architecture}\label{sec:repro}

Our reproducibility guarantees rest on three mechanisms that work
together.

\paragraph{Checksummed artifacts.}
Every result-returning function reads its reference data from the
snapshot of \cref{sec:overview}. The manifest records a SHA-256 digest
and a schema version for each of its eight files. The radius snapshot and
the certificate fixture carry their own digests, which we pin in the
version module of our package together with the commits of our research
repository from which we baked them. The snapshot check recomputes the
eight digests in about $3$~ms. Every replay of a radius certificate first
checks the fixture digest and the source commits, and refuses with the
failure state \emph{corrupt artifact} on any mismatch. We make snapshot
upgrades explicit, and every result names the snapshot against which we
computed it.

\paragraph{The result contract.}
We fix the envelope of \cref{sec:overview} by a JSON schema that we ship
with it. Our dependency-free validator rejects unknown fields, checks
that the display strings agree with the canonical expression graph, and
requires that a passing verification report contain at least one
required check with every required check passed. We treat a skipped
required check as a failure, so a missing leaf, an incomplete
certificate or an unresolved denominator cannot be reported as proven.
Our trusted-implementation registry resolves only package-owned names,
and no record may carry an import path, a formula string, or executable
code.

\paragraph{Replayable certificates.}
We replay the $306$ coefficient certificates with \cref{alg:verifycert}
and the eight reviewed radius lanes with \cref{alg:replay}. We bound both
replays with explicit limits on precision, step count, expression length
and nesting depth, and both fail closed. A mismatch in either fails our
test suite, and it fails any downstream suite that adopts the same call.

\paragraph{Testing.}
Our test suite runs on every commit. Beyond unit tests of each module, it
contains four further kinds of test. Parity tests compare our output with
the counterexample witnesses and laboratory metrics of our public
website, and with the Fekete--Szeg\H{o} artifact and radius certificate
fixture of our research repository. A documentation test runs the
documented examples and regenerates the documentation figures byte for
byte. A cross-check covers the migration from our research verifier. A
release gate installs the wheel into a fresh environment outside the
checkout, exercises the interface, confirms that the validator rejects a
verification record with one check altered, and confirms that an
untrusted implementation name cannot be resolved. We checked the listings
of this paper in the same way: a harness extracted every listing from the
manuscript source, ran it against the installed package, and compared the
printed output with the comments shown. The harness and its transcript
accompany the manuscript.

\paragraph{Provenance and citation.}
We ship a machine-readable citation file with each release, a citation
subcommand exports references in several formats, and every generator,
class, certificate and radius entry records its literature source or its
source-code locator in our research repository.

\section{Related work}\label{sec:related}

Our package sits where three software traditions meet, and we take them
in turn.

\paragraph{Symbolic algebra.}
General-purpose symbolic systems, among them SymPy~\cite{SymPy},
Mathematica, Maple and SageMath~\cite{Sage}, provide the basic
operations on which a domain-specific package must build. We depend on
SymPy for exact arithmetic and simplification, so every exact output is a
SymPy expression and works with the wider ecosystem. We deliberately
refuse formula strings and require preconstructed expressions, because
parsing a string in a computer algebra system amounts to evaluating it.

\paragraph{Rigorous numerics.}
Interval-arithmetic libraries such as mpmath~\cite{mpmath},
Arb~\cite{Arb} and INTLAB~\cite{Rump} provide certified numerical
computation. We compute the enclosures of \cref{alg:witness} with
mpmath's interval type, representing the inputs as thin intervals at
their exact binary64 values.

\paragraph{Mathematical databases.}
The OEIS~\cite{SloaneOEIS} and the LMFDB~\cite{CremonaLMFDB} have shown
how canonical identifiers, immutable snapshots and versioned interfaces
turn a scattered literature into something computable. We adopt the same
pattern for radius and coefficient problems: each entry has a stable
identifier, we checksum the underlying data, and every result carries the
identifiers of the artifacts it used.

\paragraph{Special-function libraries.}
Libraries such as SciPy's special-function module~\cite{Virtanen2020SciPy}
provide high-quality implementations of concrete transcendental
functions, and in our research repository we use them as numerical
anchors when validating our harnesses.

\paragraph{Proof assistants.}
Formal systems such as Lean~\cite{Mathlib} and Isabelle/HOL~\cite{Flyspeck}
produce proof terms checked by a small logical kernel. Our certificates
are lighter: recorded chains of identities discharged by a computer
algebra system, together with recorded classical identities and pointwise
interval enclosures. The two approaches complement each other, and we
designed the certificate format so that a future formalisation of the
Ma--Minda reductions can be combined with it; we return to this in
\cref{sec:future}.

\paragraph{Prior software in the area.}
Several recent papers on Ma--Minda classes distribute Mathematica or
Maple notebooks that evaluate the coefficient formulae for one generator.
Such notebooks are valuable records of a single computation. Our package
differs from them in three ways. We treat the generator as an object, so
a formula written once applies to every registered class. We label every
output with the kind of evidence behind it. And we ship the artifacts and
the replay needed to repeat the computation on another machine.

\paragraph{The Atlas project.}
We develop the mathematics certified by \pkg{} in our companion
paper~\cite{AtlasPaper}, where we introduce our registry, present the
sharp radius portfolio, and prove the coefficient theorems whose
certificates our package ships. In the present paper we describe the
software through which those results become independently verifiable.

\section{Conclusion}\label{sec:conclusion}

In daily practice, the extremal problems of geometric function theory are
repetitive: the same Ma--Minda reductions, the same Schur-parameter
representation and the same care about sharpness and attainment are
applied to a growing catalogue of generator-defined classes. We designed
\pkg{} to remove the repetition and keep the care. Our exact core returns
Taylor coefficients, Fekete--Szeg\H{o} constants and extremal
coefficients as exact expressions. Our verifier answers membership
questions at three named levels of evidence and keeps a screen distinct
from a proof. Our artifact snapshot ships the certificates of our
registry with digests and provenance. Our replays re-execute the recorded
proof chains of eight sharp radii and the extremal evaluations of three
hundred and six coefficient bounds, each in a fraction of a second. The
result is one interface that serves interactive research on new
generators, independent verification of published constants, and
downstream applications that need labelled numbers.

We develop the mathematics behind our package, namely the reviewed sharp
radius portfolio, the certified coefficient bounds and the catalogue of
Ma--Minda classes, in our companion paper~\cite{AtlasPaper}. The software
we describe here makes that content reproducible in the strict sense:
each constant can be recomputed or replayed in a single call, and the
certificate behind the original claim can be re-executed verbatim.

\section{Future work}\label{sec:future}

Several directions extend the present release.

\paragraph{Extending the certified radius portfolio.}
Seven of the $323$ rows with an exactly solved contact equation, and the
sine $\to$ tanh lane among the rows with a confirmed closed form, carry a
replayable chain that bounds the angular variation symbolically. For the
other rows we have confirmed the angular maximum on an $8192$-point grid.
Extending our recorded chains to these rows, so that each carries the
same symbolic bound, is the most direct way to enlarge the certified
portfolio. The $85$ unidentified rows, each carrying a sixty-digit value,
form our natural reservoir of candidate closed forms.

\paragraph{Symbolic parameters.}
Our Fekete--Szeg\H{o} function accepts a rational $\mu$. Accepting a
symbolic $\mu$ and returning the piecewise expression \eqref{eq:fs-bound}
with its transition points, in the same envelope, is a small extension
that would serve parametric studies directly.

\paragraph{Higher-order functionals.}
Our bounds for the inverse and logarithmic coefficients are sharp for all
thirty-nine classes. Our eleven second-Hankel bounds and our Zalcman
bound are certified enclosures, each with a recognised candidate constant
attained by the stored extremal. Closing these enclosures to sharp
values, and extending the catalogue to the third Hankel and higher
Toeplitz determinants, is our next step on the coefficient side.

\paragraph{Scope.}
Meromorphic starlike classes, multivalent functions and close-to-convex
families outside the Ma--Minda framework each need a modified criterion
in \cref{alg:verify} and, in the meromorphic case, a modified
normalisation. The tier machinery is independent of these details, and we
intend to add the corresponding properties in later releases.

\paragraph{Step-level scope labels.}
Our replay of \cref{alg:replay} records every step with a verdict.
Recording also whether we discharged a step by simplification, checked it
numerically, or recorded it as a classical identity would show a reader
the shape of a chain at a glance. The information is already in the
certificate text, and we can surface it in the envelope.

\paragraph{Additional generators.}
Adding a generator to our catalogue is a mechanical operation. In a
forthcoming release we will accept community-contributed definitions
through a light submission workflow, so that a newly published class can
appear in the shipped snapshot within one release cycle.

\paragraph{Proof assistants.}
We designed the certificate format for export to Lean and Isabelle/HOL.
A certificate-to-Lean bridge would let a formalised proof of the
Ma--Minda reductions and of the classical identities recorded in the
chains, once available, be combined with our chains to give
kernel-checked constants.

\paragraph*{Availability.}
We make \pkg{} available under the MIT licence from the Python Package
Index and at
\url{https://github.com/Prasanna28Devadiga/geometric-function-atlas}. The
release we describe in this paper is version \pkgversion. We maintain
continuous integration, release notes and a change log in the same
repository, and the probe, benchmark and listing-check scripts that
produced every output and timing quoted above accompany the manuscript.

\appendix

\section{Application laboratories}\label{app:labs}

Our result discipline of \cref{sec:overview} is independent of geometric
function theory. To show that it transfers, we ship two optional
laboratories in which we apply the same envelope and labels to metrics
from outside the field. We enable both with the \texttt{lab} extra, which
adds NumPy as a dependency, and we load the labs lazily, so importing our
package without the extra is unaffected.

\subsection{The cryptography laboratory}\label{sec:crypto}

Our cryptography laboratory computes the standard quality metrics of an
$8$-bit S-box for a user-supplied substitution table: nonlinearity, the
strict avalanche criterion, the bit-independence criterion, differential
uniformity, and the maximal differential and linear probabilities. It
also constructs S-boxes, by two deterministic constructions, from the
five star functions of our registry, namely the cardioid,
exponential-cardioid, nephroid, sine and quartic functions, and it
replays the immutable $435$-row leaderboard of our public website. We
state plainly that these outputs are benchmark metrics and that we make
no security claim.

\begin{lstlisting}[caption={S-box metrics with a positive and a negative anchor.},label={lst:crypto}]
from fractions import Fraction
from geometric_function_atlas.lab import AES_SBOX, IDENTITY_SBOX, sbox_metrics, website_leaderboard

aes = sbox_metrics(AES_SBOX)
print(aes["NL_min"], aes["DU"], Fraction(aes["DP"]), Fraction(aes["LP"]))   # 112 4 1/64 1/16
idn = sbox_metrics(IDENTITY_SBOX)
print(idn["NL_min"], idn["DU"], Fraction(idn["DP"]), Fraction(idn["LP"]))   # 0 256 1 1/2
print(len(website_leaderboard()))                                           # 435
\end{lstlisting}

The AES row, with nonlinearity $112$, differential uniformity $4$ and
linear probability $1/16$, is the positive anchor against which we
validate the harness at every release. The identity permutation, with
nonlinearity $0$ and differential uniformity $256$, is the negative
anchor. A drift in either fails our test suite.

\subsection{The image laboratory}\label{sec:image}

Our image laboratory pairs deterministic sample images with the standard
full-reference quality metrics: mean squared error, root mean squared
error, mean absolute error, peak signal-to-noise ratio, the Pearson
correlation, structural similarity and the gradient-magnitude similarity
deviation. It also provides a small set of finite coefficient-derived
filters used by the named functions of our website. Its outputs are
empirical, and we label them so.

\begin{lstlisting}[caption={Image metrics on an identical pair.},label={lst:image}]
from geometric_function_atlas.lab import image_metrics, sample_image

ref = sample_image(seed=0, size=32)
m = image_metrics(ref, ref)
print(m["PSNR"], m["SSIM"], m["GMSD"], m["MSE"])           # inf 1.0 0.0 0.0
\end{lstlisting}

\bibliographystyle{plain}
\bibliography{refs}

\end{document}